\documentclass[aps,pra,twocolumn,showpacs,preprintnumbers,amsmath,amssymb, superscriptaddress]{revtex4}

\usepackage{graphicx}

\begin{document}

\preprint{xxx}

\title
{
Implementation of a quantum controlled-SWAP gate with photonic circuits
}

\author{Takafumi Ono
}
\thanks{These three authors contributed equally to this work}
\affiliation{%
Research Institute for Electronic Science, Hokkaido University, Sapporo 001-0020, Japan.
}%
\affiliation{%
The Institute of Scientific and Industrial Research, Osaka University, Mihogaoka 8-1, Ibaraki, Osaka 567-0047, Japan.
}

\author{Ryo Okamoto
}
\thanks{These three authors contributed equally to this work}
\affiliation{%
Research Institute for Electronic Science, Hokkaido University, Sapporo 001-0020, Japan.
}%
\affiliation{%
The Institute of Scientific and Industrial Research, Osaka University, Mihogaoka 8-1, Ibaraki, Osaka 567-0047, Japan.
}
\affiliation{%
Kyoto University, Department of Electronic Science and Engineering, Kyoto Daigaku-Katsura, Nishikyo-ku, Kyoto 615-8510, Japan.
}

\author{Masato Tanida}
\thanks{These three authors contributed equally to this work}
\affiliation{%
Research Institute for Electronic Science, Hokkaido University, Sapporo 001-0020, Japan.
}%
\affiliation{%
The Institute of Scientific and Industrial Research, Osaka University, Mihogaoka 8-1, Ibaraki, Osaka 567-0047, Japan.
}
\author{Holger F. Hofmann}
\affiliation{%
Graduate School of Advanced Sciences of Matter, Hiroshima University, Kagamiyama 1-3-1, Higashi Hiroshima 739-8530, Japan.
}%

\author{Shigeki Takeuchi}
\email{takeuchi@kuee.kyoto-u.ac.jp}
\affiliation{%
Research Institute for Electronic Science, Hokkaido University, Sapporo 001-0020, Japan.
}%
\affiliation{%
The Institute of Scientific and Industrial Research, Osaka University, Mihogaoka 8-1, Ibaraki, Osaka 567-0047, Japan.
}
\affiliation{%
Kyoto University, Department of Electronic Science and Engineering, Kyoto Daigaku-Katsura, Nishikyo-ku, Kyoto 615-8510, Japan.
}

\date{\today}
%

\begin{abstract}
Quantum information science addresses how the processing and transmission of information are affected by uniquely quantum mechanical phenomena. Combination of two-qubit gates has been used to realize quantum circuits, however, scalability is becoming a critical problem. The use of three-qubit gates may simplify the structure of quantum circuits dramatically. Among them, the controlled-SWAP (Fredkin) gates are essential since they can be directly applied to important protocols, e.g., error correction, fingerprinting, and optimal cloning. Here we report a realization of the Fredkin gate for photonic qubits. We achieve a fidelity of 0.85 in the computational basis and an output state fidelity of 0.81 for a 3-photon Greenberger-Horne-Zeilinger state. The estimated process fidelity of 0.77 indicates that our Fredkin gate can be applied to various quantum tasks.
\end{abstract}

\maketitle
%

\section*{Introduction}

Quantum information science addresses how the storage, processing, and transmission of information are affected by uniquely quantum mechanical phenomena, such as superposition and entanglement \cite{Nielsen2000}. New technologies that harness these quantum effects are beginning to be realized in the areas of communication \cite{Gisin2007, OBrien2009}, information processing \cite{OBrien2007, Ladd2010} and precision measurement \cite{Giovannetti2004, Nagata2007a, Ono2013}. For the realization of a universal gate set, by which, in principle, any quantum information task can be realized, two-qubit gates have been demonstrated \cite{Monroe1995, O'Brien2003, Anderlini2007, DiCarlo2009, Okamoto2005} and have been used to realize small-scale quantum circuits \cite{Vandersypen2001, Martin-Lopez2012}. However, scalability is becoming a critical problem. It may therefore be helpful to consider the use of three-qubit gates \cite{Fredkin1982}, which can simplify the structure of quantum circuits dramatically \cite{Lanyon2008, Ozaydin2014}. Although both the controlled-SWAP (CSWAP) gate (also called Fredkin gate) \cite{Fredkin1982} and the controlled-controlled-NOT gate (also called Toffoli gate) \cite{Fredkin1982} are representative three-qubit gates, a crucial difference is that the Toffoli gate has two control bits and only one target bit, so that the eigenstates of the operation are local states, whereas the Fredkin gate has only one control qubit that acts on the symmetry of the two target bits, so that the eigenstates of the operation are entangled symmetric/antisymmetric states of the target qubits. For this reason, the Fredkin gates can be directly applied to many important quantum information protocols, e.g., error correction \cite{Cory1998}, fingerprinting \cite{Buhrman2001}, optimal cloning \cite{Hofmann2012}, and controlled entanglement filtering. Here we report a realization of the Fredkin gate using a photonic quantum circuit, following the theoretical proposal by Fiur\'{a}\v{s}ek \cite{Fiurasek2008}. We achieve a fidelity of 0.85 $\pm$ 0.03 for the classical truth table of CSWAP operation and an output state fidelity of 0.81 $\pm$ 0.13 for a generated 3-photon Greenberger-Horne-Zeilinger (GHZ) state. We also confirmed that the gate is capable of genuine tripartite entanglement with a quantum coherence corresponding to a visibility of 0.69 $\pm$ 0.18 for three-photon interferences. From these results, we estimate a process fidelity of 0.77 $\pm$ 0.09, which indicates that our Fredkin gate can be applied to various quantum tasks.

Since the implementation of a Fredkin gate with only single- and two-qubit gates requires five two-qubit gates and many single-qubit operations, it has not been realized yet in any qubit system \cite{Smolin1996}. For photonic qubits, the limited efficiency of the photonic quantum gates \cite{Hofmann2002} restricts the success rate of the Fredkin gate to $10^{-5}$. As has been proposed for the Toffoli gate \cite{Lanyon2008}, the use of excess degrees of freedom may ease this difficulty, but this would require the precise control of a larger Hilbert space. As an alternative approach, Fiur\'{a}\v{s}ek proposed an implementation based on multi-photon control of an interferometer \cite{Fiurasek2008}.

Fiurášek's proposal is based on single-photon interference in a Mach-Zehnder interferometer (Fig. 1A). It is well known that a single photon incident from mode A(B) is output from mode C(D) with 100\% probability when the lengths of the two optical paths are the same, i.e., when the phase is $\phi = 0$. On the other hand, when the pathlengths are different by $\lambda/2 (\phi = \pi)$, a photon incident from mode A(B) will be output from the opposite mode, D(C). Figure 1B shows an optical partial SWAP gate based on this principle \cite{Cernoch2008}, where the qubits are encoded into the polarization of photons. The target photonic qubits are input from T1in and T2in, and the gate operation is successful when the photons are output from T1out and T2out, respectively. By changing the phase $\phi$, the different gate operations, including the identity gate ($\phi =0$), the SWAP gate ($\phi =\pi$) and the $\sqrt{\text{SWAP}}$ gate ($\phi =\pi/2$), can be obtained \cite{Cernoch2008}.

\begin{figure}
\begin{center}
\includegraphics[width=9cm]{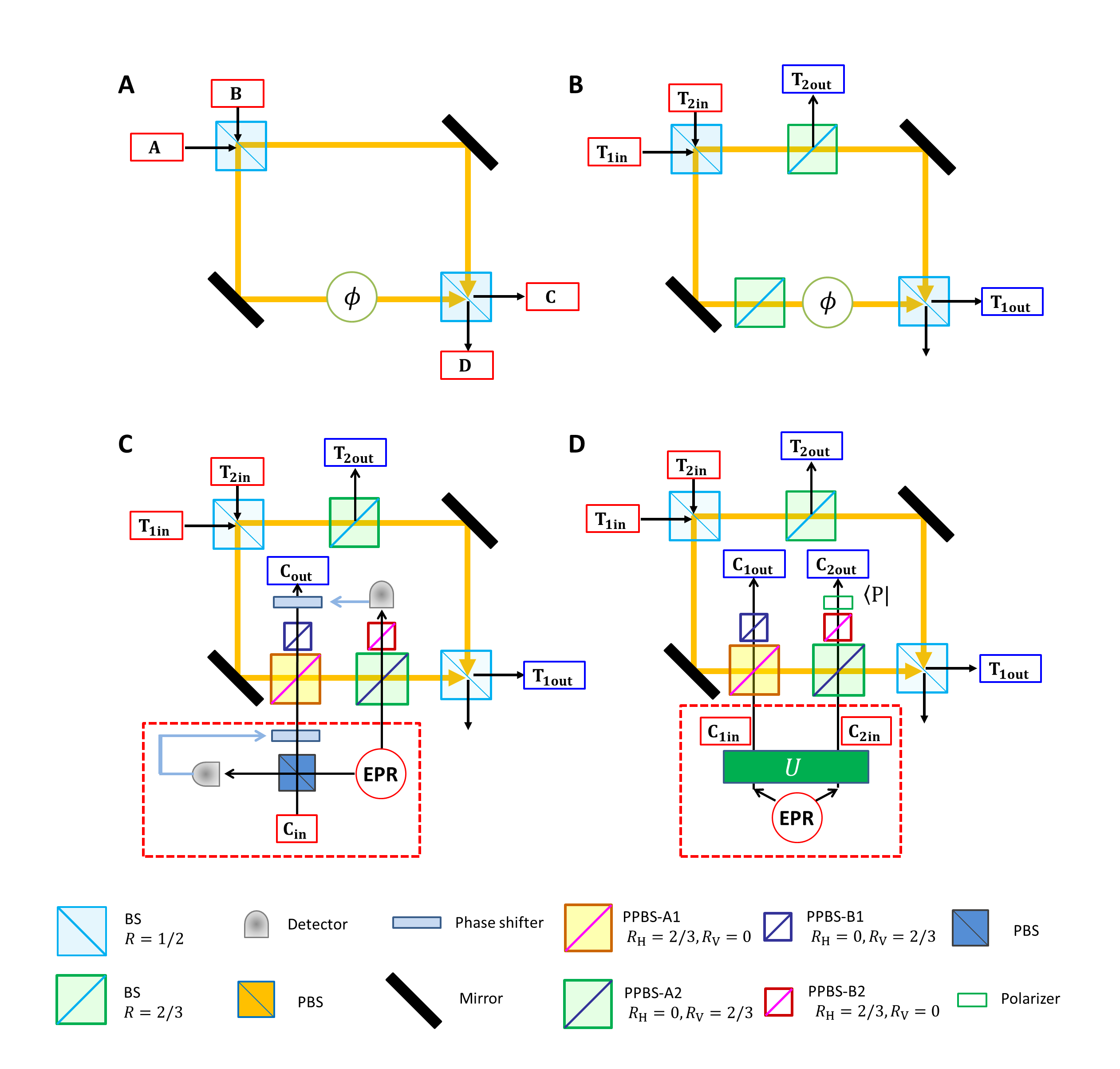}
\caption{\label{fig1}
({\bf A}) Mach-Zehnder interferometer. ({\bf B}) Partial SWAP gate. ({\bf C}) Proposal for a controlled SWAP gate by Fiur\'{a}\v{s}ek \cite{Fiurasek2008}. ({\bf D}) Simplified circuit for a controlled SWAP gate, where the parity check gate in panel ({\bf C)} is removed, and a controlled qubit is directly encoded onto an entangled photon pair by corresponding unitary operation $U$. The setup inside the squared dotted red line corresponds to that in panel ({\bf C}). The state in mode ${\rm C_{2out}}$ was projected onto plus diagonal polarization $\langle P |$ instead of the controlled phase shifter in panel ({\bf C}). PPBS: Partially polarizing beam splitter. BS: Beam splitter. PBS: Polarizing beam splitter. 
}
\end{center}
\end{figure}

A quantum CSWAP can be realized by replacing the classical phase shifter with a photonic quantum phase gate (QPG). The purpose of this QPG is for changing the operations which are controlled by the phase inside the interferometer, where the operation is identity when control q-bit is 0, that is the phase is $\phi=0$, whereas the operation is SWAP when the control q-bit is 1, that is $\phi = \pi$. Figure 1C shows the quantum CSWAP gate proposed by Fiur\'{a}\v{s}ek \cite{Fiurasek2008}. The QPG is realized by combining two optical controlled-NOT (CNOT) gates, which are based on the two-photon interferences at the partially polarizing beam splitters. The state of the control qubit is encoded into the state of the two photons incident to the CNOT gates by using an Einstein-Podolsky-Rosen (EPR) source and a quantum parity check: the encoder transforms the input state of the control photon ${\rm (\alpha | H \rangle_{C_{in}}  + \beta |V \rangle_{C_{in}} )/\sqrt{2} }$  into a state  ${\rm (\alpha | H \rangle_{C_{1in}} |V \rangle_{C_{2in}} + \beta |V \rangle_{C_{1in}} |H \rangle_{C_{2in}})/\sqrt{2} }$ with a probability of 1/2.

In this proof-of-principle demonstration, we adopted a simplified scheme (Fig. 1D), where the control qubit is directly encoded into the entangled photon pair ${\rm (\alpha | H \rangle_{C_{1in}} |V \rangle_{C_{2in}} + \beta |V \rangle_{C_{1in}} |H \rangle_{C_{2in}})/\sqrt{2} }$ generated via spontaneous parametric down-conversion and local polarization operations. The CSWAP operation is successful when the photons are simultaneously detected at the output ports(${\rm T_{1out}}$, ${\rm T_{2out}}$, ${\rm C_{1out}}$, ${\rm C_{2out}}$). To the best of our knowledge, the success probability of this operation (1/162) is the highest among the various proposals for photonic CSWAP gates. Note that for this simplified scheme, an unknown qubit cannot be used for the control qubit. However, the simplified scheme can still handle any (entangled) unknown target qubits together with a known control qubit in an arbitrary superposition state, and thus useful for many tasks, e.g. quantum finger printing and optimal cloning.

\section*{Results}
In order to demonstrate the proposed simplified CSWAP gate, the interferometer that is integrated with multiple optical quantum gates must be perfectly stabilized. As shown in Fig. 2A, we realized a highly stable photonic quantum circuit for the simplified CSWAP gate by using two key technologies: displaced-Sagnac architecture and hybrid partially polarizing beam splitters (PPBSs), which contain multiple PPBSs in one optical component. In this optical circuit, the two main optical paths of the interferometers in Fig. 1D are folded, and the two beam splitters (BSs) in Fig. 1D correspond to the blue beam splitter in Fig. 2A. In order to implement the different gate operations for each of the folded paths in the displaced-Sagnac interferometer, we used two types of hybrid PPBSs (specially ordered, OPTOQUEST Co., LTD.), each of which combines a PPBS and a mirror or BS. Since both of the folded optical paths experience the same optical components, the displaced-Sagnac architecture is intrinsically robust against disturbance. Figure 2B shows the interference fringe that occurs when a laser light is incident to ${\rm T_{1in}}$ and detected at ${\rm T_{2out}}$. The fringe maintained a very high visibility $V = 99.5 \%$ over a full week without any active stabilization.

\begin{figure*}
\begin{center}
\includegraphics*[width=15cm]{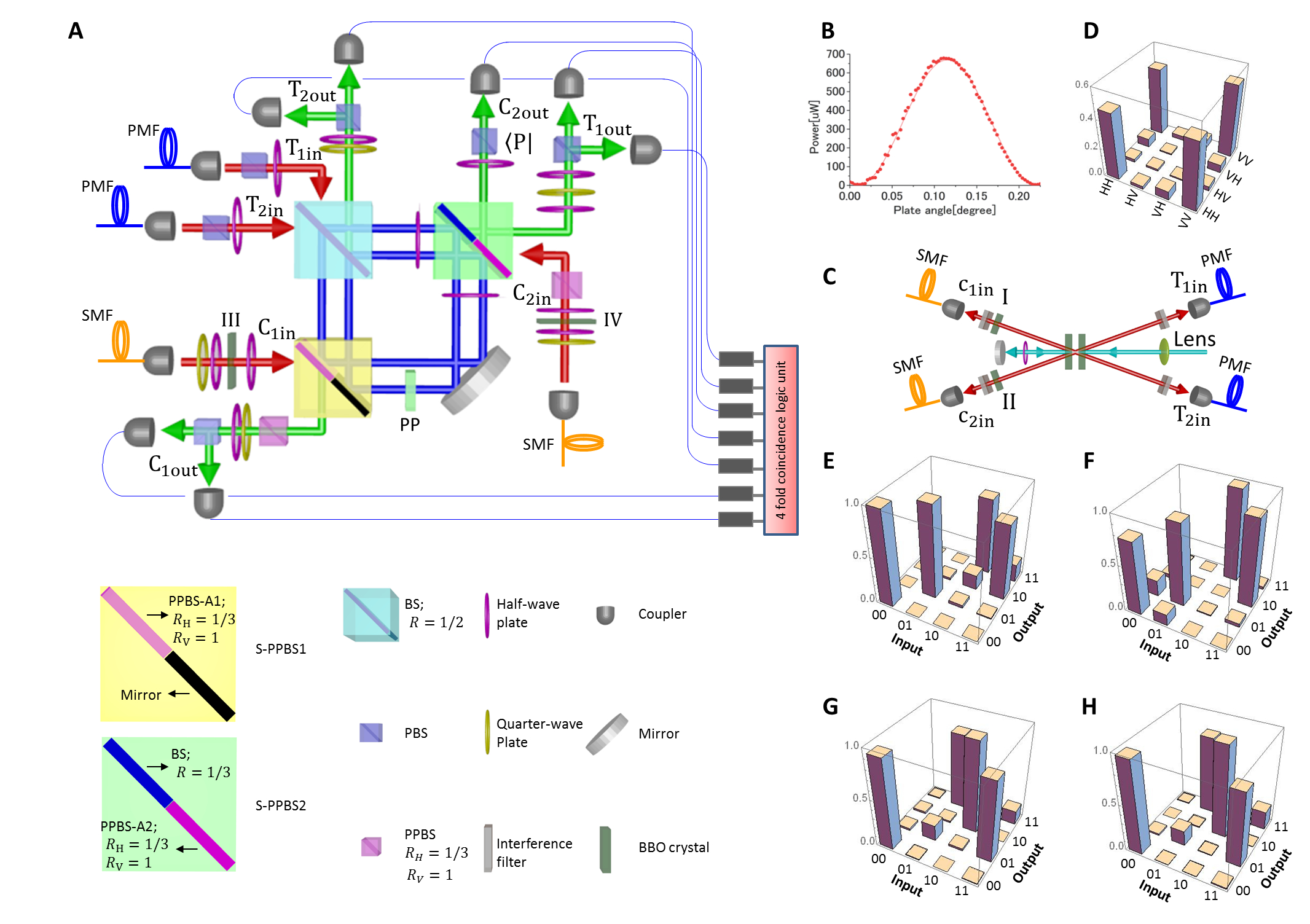}
\caption{\label{fig2}
({\bf A}) Experimental setup for controlled swap gate. H-PPBS: Hybrid partially polarizing beam splitter. The PBS reflects the vertical polarization component and transmits horizontal polarization. PPBS-As reflects vertically polarized light perfectly and reflects 1/3 of horizontally polarized light. H-PPBS1 is composed of a mirror and PPBS-A. H-PPBS2 is composed of a beam splitter and PPBS-A. ({\bf B}) Classical interference fringe of a Sagnac interferometer. While varying the phase by rotating the phase plate (PP), we measured the output power at ${\rm T_{2out}}$. ({\bf C}) Optical source. A lens (focus length = 600 mm) was placed in front of the BBO crystals to focus the beam. ({\bf D}) Absolute value of the reconstructed density matrix of the generated state measured after the circuit (after the output couplers in mode ${\rm C_{1out}}$ and ${\rm C_{2out}}$) using quantum state tomography. ({\bf E-F}) Experimental results of CNOT gate operation  in the computational basis using {\bf E} PPBS-A1 and {\bf F} PPBS-A2, where the computational basis of the gate is defined as ${\rm | 0 \rangle_C \equiv | V \rangle_C, | 1 \rangle_C \equiv | H \rangle_C }$ for the control qubit and ${\rm | 0 \rangle_T \equiv 1/\sqrt{2} (| V \rangle_T + | H \rangle_T ) }$, ${\rm | 1 \rangle_T \equiv 1/\sqrt{2} (| V \rangle_T - | H \rangle_T ) }$ for the target qubit. ({\bf G-H}) Experimental results of CNOT gate operation in the complementary basis using {\bf G} PPBS-A1 and {\bf H} PPBS-A2, where the computational basis of the gate is defined as  ${\rm | 0 \rangle_C \equiv 1/\sqrt{2} (| V \rangle_C + | H \rangle_C ) }$, $ {\rm | 1 \rangle_C \equiv 1/\sqrt{2} (| V \rangle_C - | H \rangle_C ) }$ for the control qubit and  $ {\rm | 0 \rangle_T \equiv | V \rangle_T, | 1 \rangle_T \equiv | H \rangle_T }$ for the target qubit. For the CNOT operation at PPBS-A1, input photons from ${\rm T_{1in}}$ and ${\rm C_{1in}}$ are used for the target bit and control bit, respectively. For the CNOT operation at PPBS-A2, input photons from ${\rm T_{1in}}$ and ${\rm C_{2in}}$ are used for the target bit and control bit, respectively.
}
\end{center}
\end{figure*}

For the entangled photon pairs \cite{Kwiat1999} and the two single photons \cite{Tanida2012} required for the CSWAP gate operation, we constructed the photon source shown in Fig. 2C. Ultraviolet laser pulses with a center wavelength of 390 nm from a frequency-doubled mode-locked Ti:sapphire laser (wave length: 780 nm; pulse width: 100 fs; repetition rate: 82 MHz) pumped a pair of 1.5 mm-thick Type-I $\beta$-barium borate (BBO) crystals. The pump laser light was diagonally polarized to generate the entangled photon pair ${\rm (| H \rangle_{C_{1in}} |H \rangle_{C_{2in}} + |V \rangle_{C_{1in}} |V \rangle_{C_{2in}})/\sqrt{2} }$. The pump laser was then reflected, and its polarization was changed to vertical by using a quarter-wave plate (QWP) for the generation of pairs of two single photons ${\rm | H \rangle_{T_{1in}} | H \rangle_{T_{2in}} }$ \cite{Kwiat1999}. The spatial decoherence, which is dependent on the emission angle, and the spectral-temporal decoherence, which is dependent on the group delay, are compensated by using an additional four BBO crystals (I-IV in Fig. 2) \cite{Rangarajan2009, Tokunaga2008}. The relative phase between the H and V polarizations is adjusted by a wave plate. From the result of quantum state tomography (Fig. 2D), the fidelity of the entangled state is 0.962 $\pm$ 0.002, and the entanglement concurrence is 0.941 $\pm$ 0.001, indicating that this state is almost maximally entangled. After removing the scattered pump light by using an interference filter (center wavelength 780 nm; full width at half maximum 2 nm), the two entangled photons are delivered to the circuit via single-mode fibers so that horizontally and vertically polarised photons propagate at the same speed inside the fibers, whereas the two single photons are delivered to the circuite via polarization maintaining fibers. Note that the polarizing beam splitters (PBSs) that are set just after the input couplers for ${\rm T_{1in}}$ and ${\rm T_{2in}}$ in Fig. 2A are used to increase the purity of the two target input photons. The QWPs and half-wave plates (HWPs) that are set after the input couplers for ${\rm C_{1in}}$ and ${\rm C_{2in}}$ are used to compensate the changes in polarization inside the single-mode fibers. The timing of the input photons injected to the photonic circuit is controlled by moving the positions of the fiber couplers.

The output photons in modes ${\rm T_{1out}}$, ${\rm T_{2out}}$, and ${\rm C_{1out}}$ are filtered by polarization analyzers consisting of QWPs, HWPs, and PBSs, and the photons from each of seven possible output modes are detected by single-photon counting modules (SPCM-AQ-FC, Perkin Elmer). Using a multichannel photon correlator (DPC-230, Becker $\&$ Hickl GmbH), all the fourfold coincidence events were analyzed. Note that the events where more than three photon-pairs are simultaneously generated from the BBO crystal cause errors. The contribution of such events can be reduced by decreasing the pump power, which results in lower coincidence rate. We optimized the pump power considering both the uncertainty due to the limited number of detection events and the errors caused by the excess multi-pair emission events.

Since quantum control is realized by a sequence of two CNOT operations implemented by PPBS-A1 and PPBS-A2, we first evaluated these operations separately using two sets of complementary inputs \cite{Okamoto2005}. Note that for this evaluation, the effect of the events with more than three photon-pairs were compensated using a separate measurement. Figures 2E and 2F show the results of using a CNOT operation for a computational basis for PPBS-A1 and PPBS-A2, respectively. The fidelity of the truth table of CNOT operation is 0.895 and 0.893 for PPBS-A1 and PPBS-A2, respectively. Figures 2G and 2H show the results of using a CNOT operation for a complementary basis for PPBS-A1 and PPBS-A2, respectively. The fidelity of the truth table for the complementary basis is 0.889 and 0.890 for PPBS-A1 and PPBS-A2, respectively. These results proved the high performance of the CNOT gates with two heralded single photon sources.

\begin{figure}
\begin{center}
\includegraphics*[width=9cm]{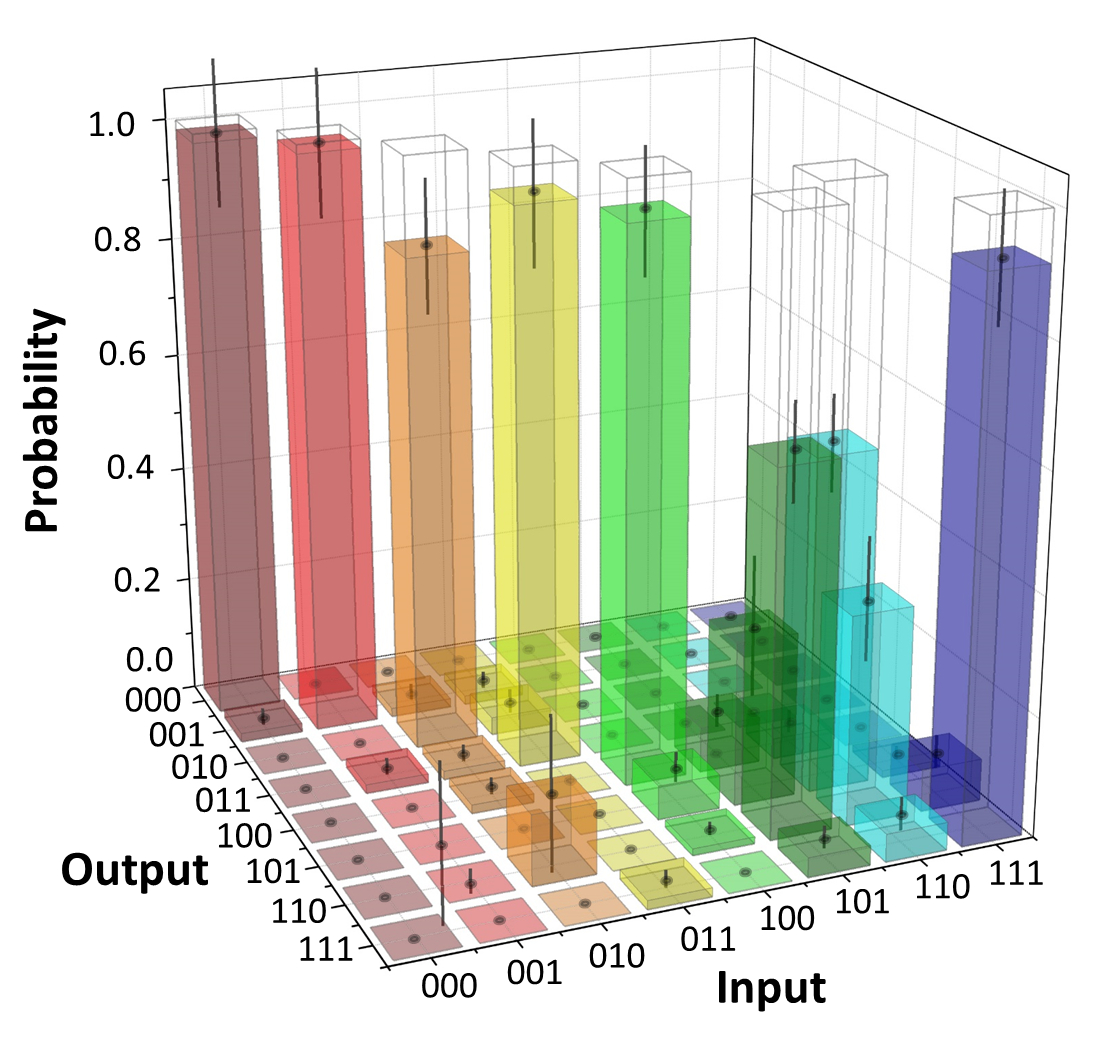}
\caption{\label{fig3}
Measurement result of our controlled swap gate in the computational basis. The input basis of the gate is defined as ${\rm | 0 \rangle_{C_{in}} \equiv | V \rangle_{C_{1in}} | V \rangle_{C_{2in}}}$, ${\rm | 1 \rangle_{C_{in}} \equiv | H \rangle_{C_{1in}} | H \rangle_{C_{2in}} }$ for the control qubit, ${\rm | 0 \rangle_{T_{1in}} \equiv | V \rangle_{T_{1in}}}$, ${\rm | 1 \rangle_{T_{1in}} \equiv | H \rangle_{t_{1in}} }$ for the target 1 qubit, and ${\rm | 0 \rangle_{T_{2in}} \equiv | V \rangle_{T_{2in}}}$, ${\rm | 1 \rangle_{T_{2in}} \equiv | H \rangle_{T_{2in}} }$ for the target 2 qubit. The output computational basis of the gate is defined as ${\rm | 0 \rangle_{C_{out}} \equiv | V \rangle_{C_{1out}}}$, ${\rm | 1 \rangle_{C_{out}} \equiv | H \rangle_{C_{1out}} }$ for the control qubit, ${\rm | 0 \rangle_{T_{1out}} \equiv | V \rangle_{T_{1out}}}$, ${\rm | 1 \rangle_{T_{1out}} \equiv |H \rangle_{T_{1out}} }$ for the target 1 qubit, and ${\rm | 0 \rangle_{T_{2out}} \equiv | V \rangle_{T_{2out}}}$, ${\rm | 1 \rangle_{T_{2out}} \equiv | H \rangle_{T_{2out}} }$ for the target 2 qubit. Measurement was implemented during 130,000 sec for each input. The events in where two pairs are generated and coupled to either ( ${\rm T_{1in}, T_{2in}}$) or ( ${\rm C_{1in}, C_{2in}}$) are compensated using separate measurements (see Methods). 
}
\end{center}
\end{figure}

Figure 3 shows the measurement result of our controlled swap gate in the computational basis. In order to convert the coincidence rates to probabilities, we normalize them with the sum of the coincidence counts obtained for each of the respective input states. In the ideal case of our implementation, the target states are swapped if and only if the controlled qubit is in the logical $| 1 \rangle $ state. In Fig. 3, the diagonal components (0.28 for $|101 \rangle \langle 101|$ and 0.35 for $|110 \rangle \langle 110|$) corresponding to the erroneous operations (non-swap for control qubit is 1) exist even though they are smaller than the correct operations (0.61 for $|110 \rangle \langle 101|$ and 0.56 for $|101 \rangle \langle 110|$), where $|110 \rangle \langle 101|$ denotes the component of 101 input and 110 output. These diagonal components increased significantly for a higher pump power, indicating that these errors are in part caused by the events where more than three photon-pairs are generated simultaneously. The classical fidelity $ F_{zzz}$ of the controlled swap operation in the computational basis, defined as the probability of obtaining the correct output averaged over all eight possible inputs, was 0.85 $\pm$ 0.03. 

\begin{figure*}
\begin{center}
\includegraphics*[width=18cm]{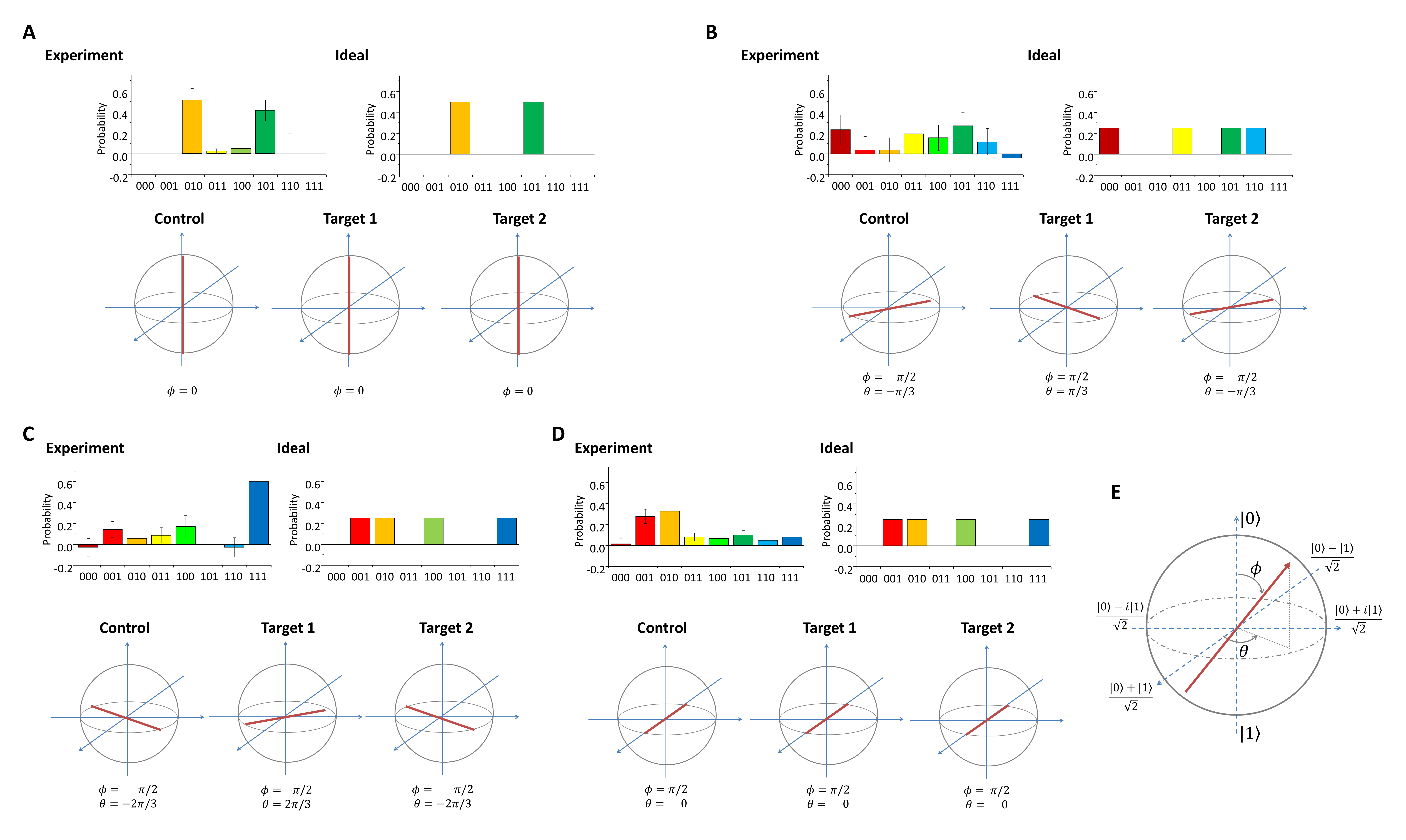}
\caption{\label{fig4.eps}
Evaluation of the quantum coherent CSWAP operation. The top of each panel ({\bf A-D}) shows the
fourfold coincidence probabilities for different measurement settings ($M_{0}$, $M_{1}$, $M_{2}$, $M_{3}$). The bottom of each panel 
shows, on a Bloch sphere, the corresponding measurement axis with its polar angle $\theta$
and azimuthal angle $\phi$, which are shown in ({\bf E}). For example, in panel {\bf A}, the top of each 
panel shows all outcomes of the measurement in the $HV$ basis for control, target 1, and 
target 2. The measurements were implemented during 320,000 sec for each measurement basis. The erroneous events where two photon-pairs are generated and coupled to either ( ${\rm T_{1in}, T_{2in}}$) or ( ${\rm C_{1in}, C_{2in}}$) are compensated using separate measurements (see Methods).
}
\end{center}
\end{figure*}

The final step is to test the genuine quantum coherent operation of the CSWAP gate. Since the count rate of $10^{-4}$ Hz is too low to perform a complete process tomography, we focused on a direct verification of the entanglement capability of the gate that could be achieved by a coherent superposition of the control qubit. Specifically, a completely local input of any two orthogonal states for the target qubits and a superposition of $(|0 \rangle + |1 \rangle)/\sqrt{2}$ for the control qubit will result in GHZ-type three-photon entanglement, where the coherence between the swap operation associated with a control qubit state of $|1 \rangle$ and the identity operation associated with $|0 \rangle$ is mapped onto a three-photon coherence in the output (see Methods). Experimentally, we generated this three-photon coherence by using a computational basis input of $|10 \rangle$ for the target photons and a superposition of $(|0 \rangle + |1 \rangle)/\sqrt{2}$  for the control photon, so that the output is given by
\begin{equation}
 |\text{GHZ} \rangle = U_{\text{gate}} (|0 \rangle + |1 \rangle) |10 \rangle/\sqrt{2} = (|010 \rangle + |101 \rangle )/\sqrt{2}.
\end{equation} 
We evaluated the three-photon coherence of this state by detecting the correlations between the three different polarization directions defined by equally spaced phase shifts between the H and V polarization of the computational basis. By averaging the parity of the measurement outcomes, we can determine the three-photon coherence of the output $\rho$,
\begin{equation}
C = 2 \text{Re} [ \langle 010| \rho |101 \rangle ].
\end{equation}
For separable states, the maximal value of C is 1/4, and for biseparable states, it is 1/2, and for ideal GHZ state, it is 1. From the experimental data shown in Fig. 4, the estimated quantum coherence of 0.69 $\pm$ 0.18 indicates the fact that under the assumption of Gaussian statistics, we have more than $85 \%$ probability for C $>$ 0.5 which can be achieved only by genuine three partite entanglement. See Methods for details.

As mentioned above, it is reasonable to assume that the entanglement capability of the quantum gate originates from the coherence between the identity operation $ |0 \rangle \langle 0| \otimes I$ and the swap operation $ |1 \rangle \langle 1 | \otimes U_{\text{swap}}$. Based on this assumption, it is possible to estimate the process fidelity by taking the average of the classical fidelity $F_{zzz}$ and the coherence C (see Methods). The estimated process fidelity, which is
defined by the normalized product trace of the process matrix and the matrix describing the ideal process, obtained from these two sets of experimental data is $F_{\text{process}} = (F_{zzz}+C)/2 = 0.77 \pm 0.09$. We are thus able to get a fairly good idea of how well the quantum process works by identifying the essential elements of process coherence directly with the non-classical coherence observed in a specific output state.

In the above analysis, C is used to test the quantumness of the gate by evaluating its entanglement capability. This is complementary to $F_{zzz}$, which tests the classical performance of the gate. $F_{\text{process}}$ is a kind of summary of both aspects, providing a single figure of merit that averages the Fidelities of the quantum features and the classical features. Importantly, the coherence C is a direct measure of the quantum coherent operation used in applications, such as cloning, fusion of entangled states, or quantum fingerprinting. Our result of $C = 0.69$ indicates that the gate can successfully operate on entangled inputs without destroying the entanglement. The setup should therefore be suitable for integration into larger quantum circuits for these tasks. 
For more information, the state fidelity for the output GHZ state is given as follows \cite{Guhne2007}:
\begin{equation}
F_{\text{GHZ}} =  \langle \text{GHZ} | \hat{\rho} | \text{GHZ} \rangle = \frac{1}{2} (M_{0} + C),
\end{equation}
where $M_0$ is the fidelity of the diagonal elements of the terms $| 010 \rangle \langle 010 |$ and $| 101 \rangle \langle 101 |$. From the result of the HV basis measurement shown in Fig. 4A, $M_0 = 0.927 \pm 0.183$. Thus, the estimated fidelity for the output GHZ state is $F_{\text{GHZ}} = 0.81 \pm 0.13$.

\section*{Discussion}

To conclude, we have successfully demonstrated a CSWAP operation with a fidelity of 0.85 $\pm$ 0.03 for the classical truth table and an output state fidelity of 0.81 $\pm$ 0.13 for a generated 3-photon GHZ state. The quantum coherence of C = 0.69 $\pm$ 0.18 indicates the fact that under the assumption of Gaussian statistics, we have more than 85\% probability for C $>$ 0.5 which can be achieved only by genuine three partite entanglement. The estimated process fidelity of 0.77 $\pm$ 0.09 also indicates that our CSWAP gate is suitable for integration into larger quantum circuits. For the erroneous operation of the gate (23 $\%$), our rough estimation suggests that eighty percent of such errors are due to the imperfection of the current photon source which sometimes emits more than three photon-pairs (up to about 10 $\%$) simultaneously and has non-zero timing jitter due to group velocity mismatch (See Methods for details). One of the current challenges in the community is to improve the single photon sources with smaller excess photon components in the output. Examples are heralded single photon sources using multiple SPDC sources in parallel \cite{Ma2011, Mendoza2016}, and single photon sources using quantum dots \cite{Ding2016, Somaschi2016}. The current experimental demonstration also highlights the strong demand for such sources.

Although a measurement based approach is promising for future linear optics quantum computation in terms of efficiency \cite{Gimeno-Segovia2015}, photonic quantum circuits combining quantum gates are also useful for various applications or tasks for quantum information processing and quantum communication. The eigenstates of a multi-photon gate determine which states can be measured and distinguished by observing local changes in one of the photons. Since the eigenstates of the CSWAP gate are entangled symmetric/antisymmetric states of the target qubits, the CSWAP is especially useful for all implementations that require a QND measurement of entanglement. Next challenge would be the realization of the non-simplified scheme (Fig. 1C) for an unknown control qubit embedded in a polarization of a single photon. Although the probability of the success of the gate is the same with the simplified scheme (1/162), one needs to precisely control 5 photons (3 single photon inputs plus one ancillary entangled photon pair.) The demonstrated controlled-SWAP gate is successful only when the resultant single photons are output from each of the output modes one by one, which may limit its potential uses within larger quantum circuits without technically demanding non-demolition photon number measurements. A fully heralded controlled-SWAP gate using linear optics is also possible but would require a total of nine single photons \cite{Fiurasek2008}. For such experiments, efficient single photon sources will be important \cite{Ding2016}. Note also that since unknown photonic qubits encoded in polarization can be used as two target bits and the resultant state is output as three single-photon qubits, our scheme is directly applicable for interesting tasks including quantum fingerprinting \cite{Buhrman2001}, optimal cloning \cite{Hofmann2012}, and can be used as a built-in component in photonic quantum circuits for other protocols. An alternative realization of an optical quantum controlled SWAP operation has recently been reported by Patel et al \cite{Patel2016}.

\section*{Methods}
\subsection*{Measurement results for three-photon coherence}
The essential difference between a quantum controlled swap and a classical controlled swap is that the quantum coherent superposition of the swap operation is conditioned by a control bit in the 1 state, and the identity operation is conditioned by a control bit in the 0 state, as expressed by the unitary operator of the ideal operation,
\begin{equation}
\hat{U}_{\mathrm{gate}} = \mid 0 \rangle \langle 0 \mid \otimes \hat{I} +  \mid 1 \rangle \langle 1 \mid \otimes \hat{U}_{\mathrm{swap}}.  
\end{equation}
The amount of coherence in the actual operation can be quantified in terms of the visibility of the quantum interference between the identity and the swap operation. Since it seems unlikely that random gate errors will result in the same kind of coherence, the coherence of the gate operation should correspond to the three-photon coherence generated by a coherent superposition of 0 and 1 in the control bit input,
\begin{equation}
\hat{U}_{\mathrm{gate}} \frac{1}{\sqrt{2}} (\mid 0 \rangle + \mid 1 \rangle) \mid 0 1 \rangle
= \frac{1}{\sqrt{2}} (\mid 010 \rangle + \mid 101 \rangle).
\end{equation}
The density matrix element describing the coherence between 010 and 101 originates from the process matrix element that describes the coherence between the identity operation and the swap operation, so the quantum coherence of the gate can be evaluated by experimentally determining the three-photon coherence of the output, given by
\begin{equation}
C = 2 \mbox{Re}[\langle 010 \mid \hat{\rho} \mid 101 \rangle].
\end{equation}
For separable states, the three-photon coherence C is limited by the maximal local coherence of $\langle0|\hat{\rho}_{\text{local}} |1\rangle \leq 1/2$. For completely separable states, the maximal value of C is therefore$2 \times (1/2)^3 = 1/4$. For biseparable states, the limit is given by the product of the maximal local coherence and the maximal two-photon coherence, $C \leq 2 \times (1/2)^2 = 1/2$. Therefore, three-photon coherences C larger than 0.5 indicate genuine tripartite entanglement.

Experimentally, the three photon coherence of the state results in strong correlations between the Bloch vector components in the equatorial plane. These correlations are easily observed as correlations between different polarization directions. Importantly, maximal correlations can be obtained whenever the directions of the Bloch vector components satisfy a specific relation between their angles in the XY-plane of the Bloch sphere. If we define $\hat{S}(\phi) = \hat{X} \cos \phi + \hat{Y} \sin \phi$, we find that $\hat{S}(\phi_1) \otimes \hat{S}(\phi_2) \otimes \hat{S}(\phi_3)$  is exactly equal to one whenever $\phi_1-\phi_2+\phi_3$ is an even multiple of $\pi$, or minus one if it is an odd multiple. This provides us with a particularly strong experimental criterion for three partite entanglement, since a rotation of each polarization by an angle of $\pi/3$ results in a sign flip for the correlation that is only obtained for local polarizations along $\phi=0$ if the angle of rotation exceeds $\pi/2$. 

In the following, we measure the three photon correlation for rotation angles of $0$, $\pi/3$ and $2\pi/3$, given by $\langle\hat{M_1}\rangle,\langle\hat{M_2}\rangle,\langle\hat{M_3}\rangle$ 
\begin{eqnarray}
\hat{M_1} &=& - \hat{S}(-\pi/3) \otimes \hat{S}(\pi/3) \otimes \hat{S}(-\pi/3),
\nonumber \\
\hat{M_2} &=& \hat{S}(-2\pi/3) \otimes \hat{S}(2\pi/3) \otimes \hat{S}(-2\pi/3),
\nonumber \\
\hat{M_3} &=& \hat{S}(0) \otimes \hat{S}(0) \otimes \hat{S}(0).
\end{eqnarray}
Note that $\langle\hat{M_i}\rangle$ are the measure for the correlation taking a value between -1 to 1. These three correlations can be used to reconstruct the total three photon coherence,
\begin{equation}
C = \frac{1}{3} (\langle \hat{M_1} \rangle + \langle \hat{M_2} \rangle + \langle \hat{M_3} \rangle ),
\end{equation}
Thus, genuine three partite entanglement can be demonstrated if the average value of $\langle\hat{M_i}\rangle$ for the three polarization directions separated by angles of $\pi/3$ on the Bloch sphere is greater than 1/2. In order to derive eq. (8) from eq. (6), one may use the relations $| 010 \rangle \langle 101 | = (\hat{X} + i\hat{Y}) (\hat{X} - i\hat{Y}) (\hat{X} + i\hat{Y})/8$ and $| 010 \rangle \langle 101 | + | 101 \rangle \langle 010 | = (\hat{X}\hat{X}\hat{X} + \hat{X}\hat{Y}\hat{Y} - \hat{Y}\hat{X}\hat{Y} + \hat{Y}\hat{Y}\hat{X} )/4$.  In $\hat{M}_1$ and $\hat{M}_2$, the coefficients for $\hat{X}$ are 1/2 and the coefficients for $\hat{Y}$ are $\sqrt{3}/2$, and for even numbers of $\hat{X}$ the sign is opposite. As a result, we have $\hat{M}_1 + \hat{M}_2 = - \hat{X}\hat{X}\hat{X}/4 +3\hat{X}\hat{Y}\hat{Y}/4 - 3\hat{Y}\hat{X}\hat{Y}/4 + 3\hat{Y}\hat{Y}\hat{X}/4$. Note also that $\hat{M}_3 = \hat{X}\hat{X}\hat{X}$.

From the experimental results shown in Fig. 4, we obtained values of $\langle \hat{M}_1 \rangle=0.615 \pm 0.401$, $\langle \hat{M}_2 \rangle=0.943 \pm 0.319$, and $\langle \hat{M}_3 \rangle=0.508 \pm 0.152$. From these values C = 0.69 $\pm$ 0.18. The average value of each operator was calculated in such a way that, for example, $\langle \hat{M}_2 \rangle = - P(000) + P(001) + P(010) - P(011) + P(100) - P(101) - P(110) + P(111)$, where P($ijk$) is the probability that the logical value of $ijk$ is detected in the measurement basis shown in Fig. 4C.

It may be worth noting that the high result for $\langle \hat{M}_2 \rangle$ originates mostly from an unexpected dominance of 111 outcomes, indicating a contribution from local polarizations. However, the correlations $\langle \hat{M}_1 \rangle$ and $\langle \hat{M}_3 \rangle$ still exceed 1/2, which is a significant indication of the special correlation only observable for genuine three-partite entanglement. For example, for a separable state with $\langle \hat{M}_2 \rangle = 1$, the values of $\langle \hat{M}_1 \rangle$ and $\langle \hat{M}_3 \rangle$ would be -1/8 each.

\subsection*{Estimated process fidelity}
The critical problem in the realization of the quantum process is to maintain coherence between the identity operation conditioned by a control qubit input of zero and the swap operation conditioned by a control qubit input of one. We can therefore describe the process as a quantum superposition of these two processes, using a process matrix where the coherence between the two processes is given by the corresponding off-diagonal elements. In this representation of the process matrix, the classical fidelity $F_{zzz}$ is given by the average of the diagonal elements and the coherence C is given by the off-diagonal elements. Here we assume that no other process matrix elements contribute to $F_{zzz}$ and C (in other words, the errors observed in the classical operation do not increase the coherence C observed in the quantum operation). More specifically, the errors observed in the classical operation do not result in an increase in the quantum coherence C, so that the coherence C originates only from the intended process and not from other sources. Since it seems unlikely that the indended three photon coherence arise from accidental errors, this seems to be a reasonable assumption. As the result, we can estimate the process fidelity $F_{\text{process}}$, which is defined by the normalized product trace of the process matrix and the matrix describing the ideal process as follows:
\begin{equation}
F_{\mathrm{process}} = \frac{1}{2}(F_{zzz} + C).
\end{equation}
In order to derive this relation, note that the process is a superposition of $|0\rangle \langle 0 | \otimes \hat{I}$ and $|1\rangle \langle 1 | \otimes \hat{U}_{\text{swap}}$  as shown in eq. (4) and the coherence C is a direct measure of the coherence between these two operations. In the process matrix, these can be represented by two diagonal elements and two off-diagonal elements, where $F_{zzz}$ is the product trace of the real process and the diagonal elements, and C is the product trace of the real process and the off-diagonal elements.
Based on these assumptions, our experimental results correspond to an estimated process fidelity of 0.77 $\pm$ 0.09.

\subsection*{Subtracted events}
In order to evaluate the performance of the gate, we need to know the four-fold coincidence counts caused by the case where a pair of photon is generated from each of the two parametric down converters (PDCs). However, there are cases where two pairs are generated from only one of the PDCs. We estimated the coincidence events caused by the latter events by performing separate measurements where the output of one of the PDCs was closed and subtracted the events observed in these measurements from the total four-fold detection. The process fidelity is degraded to 0.54 $\pm$ 0.03 if we do not perform this subtraction.
 
\subsection*{Experimental behavior of the custom hybrid optic elements}
The measured reflectance for H and V polarization of the components of the hybrid optics are as follows. For S-PPBS1, PPBS-A1 $\mathrm{R_H}$ = 34\%, $\mathrm{R_V}$ = 98\%, Mirror $\mathrm{R_H}$ = 99\%, $\mathrm{R_V}$ = 100\%. For S-PPBS2, PPBS-A2 $\mathrm{R_H}$ = 36\%, $\mathrm{R_V}$ = 98\%, BS $\mathrm{R_H}$ = 34\%, $\mathrm{R_V}$ = 38\%.

\subsection*{The effect of possible error sources on the performance of the circuit}
Here we roughly estimate the effect of error sources on the non-unity quantum process fidelity of 77 $\pm$ 9\%. 
The average of the observed visibilities of the two photon interference between the photons from different sources at PPBSs are 86.2\%, where the pump laser power was enough attenuated to suppress the multi-pair emission events and the actual polarization-dependent reflectivity of the PPBSs (shown above) are taken into account. This degradation in the visibility is due to the spatio-temporal mode mismatch of the photons \cite{Tanida2012}. Based on the analysis similar to Ref. \cite{Nagata2010}, we have found that the quantum process fidelity degrades about 7\%. The same analysis also suggests that the difference of the measured transmittances and reflectance of the PPBSs (shown above) from ideal ones may have caused degradation in the process fidelity by 0.2\%. The imperfection of the entangled photons with the state fidelity of 0.962 may have caused about 2\% degradation in the process fidelity. The sum of the estimated errors due to the above mentioned causes are about 9.2\%.
We strongly conjecture that the rest of the errors ($\sim$ 14\%) in the total error ($\sim$ 23\%) is mainly due to the events where three pairs of photons are emitted simultaneously from the source. By decreasing the pump power, the probability of having three-pair-emission can be suppressed but the data accumulation time has to be increased. As a compromise between these factors, we used a pump power with which the amount of four-fold coincidence events due to the three-pair-emission is about 10\%, which was estimated by a separate experiment. A more complete and detailed analysis of errors in the photonic network \cite{Tichy2015}will be one of the important future research topics, but out of the scope of this paper.

\section*{Acknowledgements}
This work was supported in part by Quantum Cybernetics of JSPS, FIRST of JSPS, a Grant-in-Aid from JSPS, JST-CREST, Special Coordination Funds for Promoting Science and Technology, Research Foundation for Opto-Science and Technology, and the GCOE program.

\clearpage

\end{document}